# Spectrum Aggregation for 6G: Lessons from 5G Carrier Aggregation and Dual Connectivity

Xingqin Lin
NVIDIA
Email: xingqinl@nvidia.com

***Abstract*—Spectrum aggregation has been a key enabler of LTE and 5G capacity growth and will remain essential in 6G as networks combine fragmented spectrum across low bands, existing mid bands, new upper-mid/centimetric bands, and millimeter wave bands. This article examines how 5G carrier aggregation (CA) and dual connectivity (DC) inform the design of 6G spectrum aggregation. The central thesis is that CA is the necessary carrier-level scaling mechanism, while DC is a higher-layer cell-group architecture that should be used only when a second cell group is genuinely needed. We argue that CA provides a cleaner and more scalable foundation for multi-band operation in standalone 6G. Building on lessons from 5G, we advocate enhanced CA as the preferred 6G spectrum aggregation framework and point out the corresponding key enhancement directions.**

## I. INTRODUCTION

Spectrum aggregation has been one of the most important techniques in modern cellular systems [1]. Starting from long-term evolution (LTE) carrier aggregation (CA) and continuing into 5G new radio (NR) developed by the 3rd generation partnership project (3GPP), the ability to combine multiple spectrum resources has enabled operators to increase peak rates, improve cell-edge throughput, and make practical use of fragmented spectrum holdings [2]. Spectrum aggregation will remain a foundational capability for 6G [3][4]. The reason is not that 6G will necessarily start with a radically different spectrum landscape, but that operators will continue to operate fragmented spectrum assets across frequencies [5]. Compared to 5G, the near-term 6G spectrum delta will likely involve additional upper-mid-band or centimetric opportunities [6].

In 6G, sub-1 GHz bands will remain essential for nationwide coverage, deep indoor reach, rural connectivity, and low-power internet of things (IoT). Existing mid bands below 6 GHz will continue to provide the main wide-area capacity layer, benefiting from mature massive multiple-input multiple-output (MIMO) deployments and broad device support. Upper-mid bands (such as the upper 6 GHz) and 7-15 GHz range are emerging as the most important new 6G candidates because they offer more bandwidth than today's mid bands while still having propagation characteristics that may support macro-cellular deployment with advanced antenna arrays [7]. Therefore, world radiocommunication conference 2027 (WRC-27) studies cover bands such as 4.4-4.8 GHz, 7.125-8.4 GHz, and 14.8-15.35 GHz, while many countries are also considering the upper 6 GHz band for international mobile telecommunications (IMT) usage [8]. At higher frequencies, millimeter wave (mmWave) will provide localized extreme capacity rather than wide-area coverage. The potential sub-THz bands (not expected to be part of day-1 6G), such as W-band and D-band, may provide extremely large bandwidths and potentially Tbps-class links, but with limited coverage and mobility. These higher frequency bands are therefore more suitable for niche scenarios such as extreme local connectivity, device-to-device links, wireless data-center connectivity, and fronthaul/backhaul. The implication is that no single spectrum layer can satisfy all 6G requirements [9]. Therefore, 6G must be designed from the beginning to aggregate spectrum across layers in different frequency ranges (FRs), as summarized in Table 1.

| Spectrum layer | Role in 6G | Strength | Limitation | Aggregation implication |
|---|---|---|---|---|
| **Sub-1 GHz** | Coverage, rural, deep indoor, low-power IoT | Excellent propagation | Limited bandwidth | Anchor mobility and reliability |
| **1-6 GHz mid band** | Wide-area macro capacity | Strong ecosystem, massive MIMO maturity | Insufficient alone for 6G capacity | Aggregate with upper-mid bands for capacity scaling |
| **Upper 6 GHz / 7-15 GHz** | New wide-area 6G capacity layer | Wider channels with more macro relevance than mmWave | More challenging propagation and coexistence | Candidate main 6G expansion layer |
| **24-52.6 GHz mmWave** | Dense urban, venues, fixed wireless access | Very large bandwidth | Blockage and limited coverage | Opportunistic capacity boost |
| **Above 52.6 GHz mmWave and sub-THz*** | Extreme local capacity, Xhaul, fixed links | Very wide bandwidth, extreme rates | Limited mobility and maturity | Specialized high-capacity layer |
| *Note: This spectrum layer is not expected to be part of day-1 6G. | | | | |

**Table 1: Comparative summary of 6G spectrum layers and their aggregation implications.**

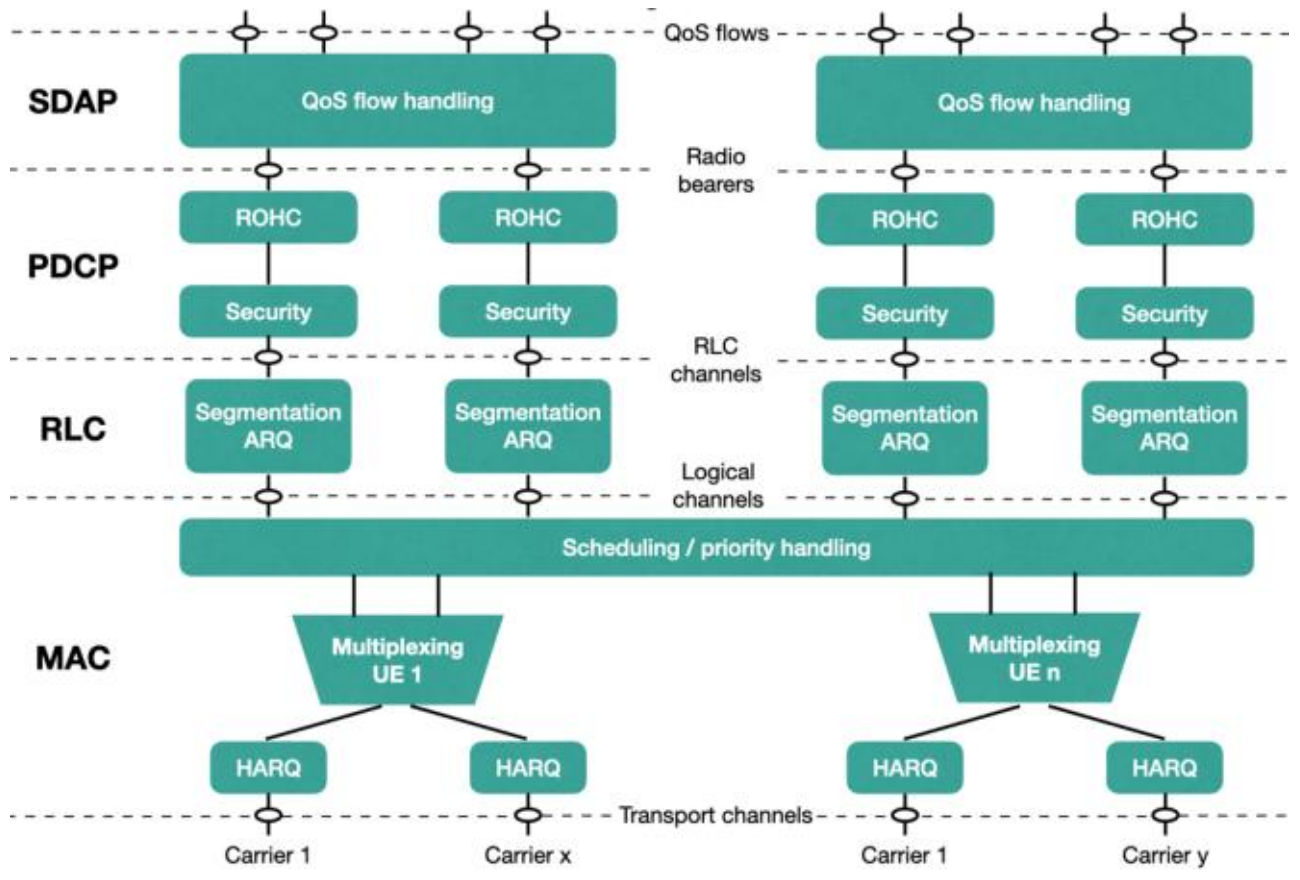

**Figure 1: An illustration of radio protocol architecture for 5G downlink CA [11].**

5G already provides a useful reference point for spectrum aggregation, but the relationship between CA and dual connectivity (DC) is sometimes misunderstood. CA is the carrier-level aggregation mechanism [10], as illustrated in Figure 1. In the 3GPP protocol model, CA is commonly described as multiple serving cells controlled by one MAC entity and a coordinated scheduler. This description is useful for specification purposes, but it should not be interpreted as a restriction that all aggregated carriers must originate from one physical cell site or from one narrowly defined scheduler implementation. In commercial 5G deployments, 5G CA has been used not only for co-sited carriers but also for non-collocated FR1 carrier deployments when the required timing, control, and scheduling dependencies can be accommodated.

DC, by contrast, is a cell-group architecture [12], as illustrated in Figure 2. Its most important commercial 5G forms have been EN-DC, where LTE provides the master cell group (MCG) and NR provides the secondary cell group (SCG), and FR1-FR2 NR-DC, where an NR FR1 layer and an NR FR2 layer are operated as separate cell groups. These deployments may be co-located or non-collocated. Importantly, DC does not replace CA. In a DC configuration, the MCG may itself use CA, and the SCG may also use CA. Thus, EN-DC can involve LTE CA on the LTE side and NR CA on the NR side; similarly, NR-DC can involve CA within the FR1 cell group, within the FR2 cell group, or both. CA and DC are therefore hierarchical rather than equivalent alternatives: CA pools carriers within a cell group, while DC introduces a second cell group when aggregation spans sufficiently independent nodes, protocol stacks, or scheduling domains.

This historical distinction matters for 6G. EN-DC was introduced largely because LTE coverage, mobility, and evolved packet core (EPC) based operation were already widely deployed when NR capacity layers were added. FR1-FR2 NR-DC has similarly been useful for rapid deployment of high-capacity layers and for deployments where the 5G CA framework imposed inconvenient cross-carrier dependencies. This article draws lessons from 5G CA and DC and discusses how these lessons can guide 6G spectrum aggregation design. The central thesis is that CA is the necessary carrier-level scaling mechanism, while DC is a higher-layer cell-group architecture that should be used only when a second cell group is genuinely needed. For pure spectrum aggregation, 6G should aim to remove some of the historical reasons that forced operators toward DC and should instead use an enhanced CA framework as the scalable baseline.

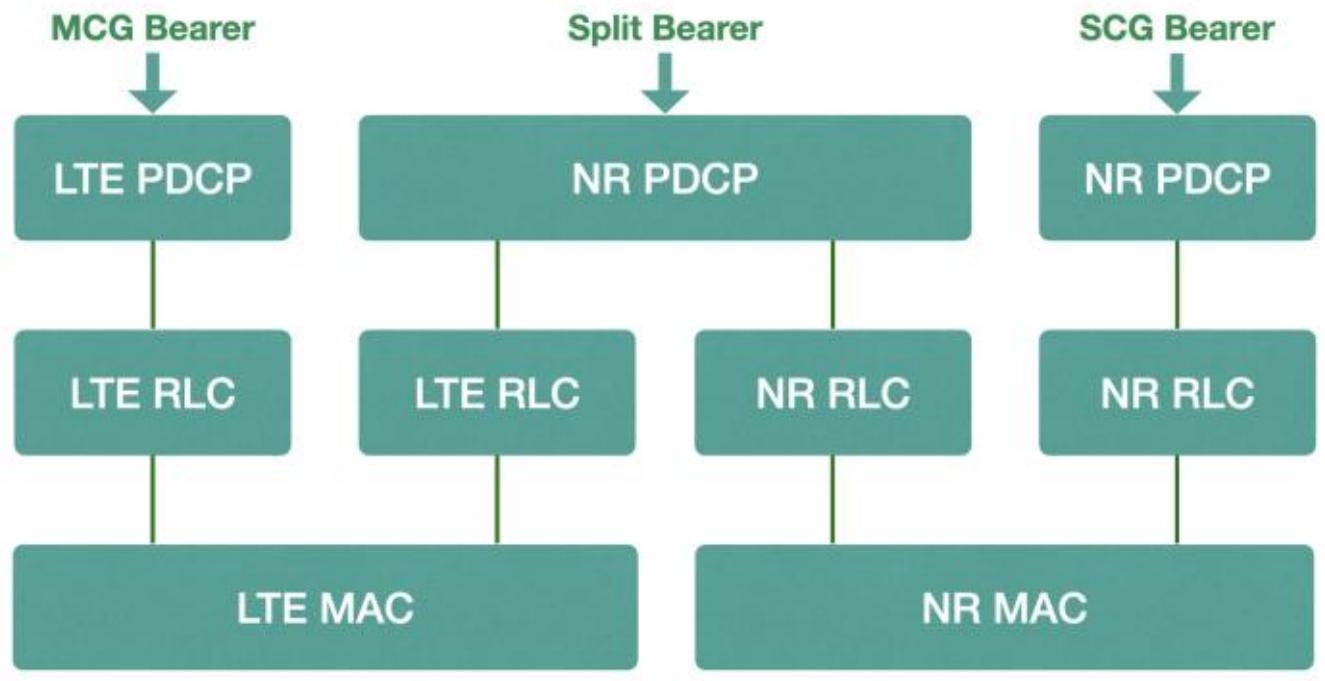

**Figure 2: An illustration of radio protocol architecture for EN-DC [13].**

## II. LESSONS FROM 5G CA AND DC

A key 5G lesson is that CA and DC are not peer alternatives. CA is a scalable carrier-level mechanism. A user equipment (UE) can receive data over multiple downlink carriers while relying on a much smaller number of uplink carriers, and in many cases on one robust uplink carrier. This asymmetry is essential for higher frequency spectrum aggregation, where a downlink carrier may provide useful capacity even when the corresponding uplink carrier has limited coverage. DC does not scale in the same way. DC introduces a second cell group, and each cell group needs its own primary cell (PCell) or primary secondary cell (PSCell) and a sustainable uplink carrier. Therefore, DC is not the mechanism that scales the number of downlink carriers. Instead, when many downlink carriers are needed in DC, CA is used within the MCG, within the SCG, or both. The resulting architecture is hierarchical: CA sits at the lower carrier-aggregation layer, while DC sits above it as a cell-group separation mechanism. This hierarchy has an important implication for 6G. If the reason for introducing a second cell group is different radio access technology (RAT) operation, legacy interworking, or very non-ideal backhaul, then DC-like operation may still be justified. However, if the motivation is only to aggregate fragmented spectrum across multiple carriers, then introducing a second cell group is unnecessary architectural hierarchy.

A second lesson is that CA and DC entered commercial 5G with different roles. In early non-standalone 5G deployments, EN-DC was the practical foundation for rapid 5G rollout. LTE provided wide-area coverage, mature mobility, and control-plane anchoring, while NR provided high-throughput capacity on newly deployed mid-band or mmWave carriers. This allowed operators to introduce 5G capacity without immediately deploying a fully standalone 5G core (5GC) and complete NR coverage layer. From a commercial perspective, DC was not only a throughput feature but also an architectural bridge from LTE to NR. By contrast, CA becomes increasingly important as NR deployments mature. Once operators deploy standalone NR and multiple NR carriers, the value of CA grows because it allows fragmented spectrum assets to be pooled efficiently. For example, an operator may combine a low-band NR carrier for coverage with a mid-band TDD carrier for capacity, or aggregate multiple mid-band carriers to increase peak and cell-edge throughput. In mature 5G networks, CA is often the cleaner mechanism for bandwidth scaling when carriers can be managed within one carrier-level aggregation framework, including co-sited deployments, shared-baseband deployments, and selected non-collocated deployments with manageable timing and control dependencies.

A third lesson is that CA is generally more scheduler-efficient, while DC is more deployment-flexible. Since CA is typically controlled within one scheduling framework, the scheduler can jointly consider channel quality, buffer status, hybrid automatic repeat request (HARQ) state, traffic priority, and carrier load across multiple serving cells. This makes CA attractive for maximizing spectral efficiency and reducing coordination delay. DC, however, allows the network to combine resources that may not be controlled by the same scheduler, such as LTE and NR nodes, macro and small-cell nodes, or geographically separated NR nodes. This makes DC more flexible, but the flexibility comes at the cost of more complex bearer management, inter-node coordination, and packet reordering. However, this deployment flexibility translates into user-perceived gain only when the traffic volume is large enough to amortize SCG addition, bearer reconfiguration, inter-node coordination, and possible packet data convergence protocol (PDCP) reordering. For short or bursty transfers, the data may already be delivered through the MCG before the SCG becomes usable. CA also has activation delay, but activation of preconfigured or dormant SCells can often be faster and lighter than adding a second cell group.

A fourth lesson is that control overhead and UE power consumption become key design constraints, while downlink aggregation is easier to exploit than uplink aggregation. Aggregating more carriers or maintaining two cell groups increases physical downlink control channel (PDCCH) monitoring, channel state information (CSI) measurement and reporting, HARQ feedback, radio frequency (RF) chain activity, and baseband processing. Commercially, the objective is not to keep every carrier or cell group active all the time, but to activate additional resources only when the expected traffic and radio conditions justify the cost. In the downlink, the network transmits from infrastructure nodes with fewer power and RF constraints. In the uplink, the UE must share limited transmit power across carriers or nodes, and simultaneous transmission may be limited by RF architecture, thermal constraints, exposure regulations, and intermodulation issues. Therefore, many commercial deployments initially emphasize downlink CA or downlink DC gains.

Uplink CA and uplink DC are valuable for high-throughput uplink services, but they are more constrained and require careful power-control and scheduling design. 5G already includes relevant building blocks, including uplink power control, power headroom reporting, UE capability signaling, prioritization rules when simultaneous uplink transmissions exceed the UE power budget, and high-power UE support such as 26 dBm Power Class 2 operation for selected bands and carrier combinations. In 6G, it would be desirable to make higher-power UE operation more mainstream for more bands and carrier combinations. Note that uplink power sharing in DC is harder than in CA because the MCG and SCG schedulers may make uplink grants with incomplete or delayed knowledge of each other's power usage. Therefore, when the carriers can be managed within one scheduling domain, CA provides a cleaner basis for protecting critical uplink control while using uplink data carriers opportunistically.

A fifth lesson is that DC is highly sensitive to transport quality. In split-bearer operation, packets may traverse both the master node and the secondary node paths. If the two paths have different latency, congestion, or scheduling behavior, PDCP reordering delay may reduce the expected gain. Similarly, if the inter-node interface is slow or congested, the secondary node may not be used efficiently. This means that DC performance is not only determined by radio bandwidth but also depends on backhaul latency, inter-node flow control, vendor interoperability, scheduler coordination, and radio resource management (RRM) policies. A related lesson is that DC, especially split-bearer operation, came with substantial 3GPP specification complexity. A split-bearer operation requires bearer-type selection, PDCP routing over multiple radio link control (RLC)/medium access control (MAC) paths, inter-node flow control, PDCP reordering, duplication control, etc., resulting in significant complexity in 3GPP specifications.

Finally, while EN-DC was commercially effective for fast 5G launch, it also made DC carry two roles at once: a useful performance feature and a temporary architectural bridge. As a result, operators had to manage LTE/NR interworking, EPC to 5GC migration, non-standalone and standalone device behaviors, separate optimization paths for EN-DC and NR-only operation, and delayed use of native 5GC capabilities [14]. By contrast, CA is a cleaner long-term spectrum-scaling tool because it aggregates carriers inside a native NR system under tighter scheduler control, rather than relying on another legacy node as the control anchor. The 6G lesson is therefore that DC should not be used as the default migration shortcut or default aggregation mechanism. 6G should start from a standalone architecture with 6G radio access available across the needed spectrum bands, so that spectrum aggregation can be designed natively through CA-like multi-band operation. Nonetheless, DC should not be characterized only as a non-standalone migration shortcut. DC also has standalone value. NR-DC has been used to combine FR1 and FR2 layers when the FR1 macro layer provides broad coverage and one or more FR2 cells provide localized capacity. Looking toward 6G, some operators may also have deployment reasons to retain 5G nodes as secondary legs under a standalone 6G system, because existing 5G coverage and capacity layers will not be replaced everywhere at once.

## III. ENHANCED CA AS THE PREFERRED BASELINE

A key 6G design question is how many historical motivations for DC can be eliminated by enhancing CA. If two carriers or carrier groups can be handled through CA with acceptable timing, feedback, control, and scheduling behavior, then a separate DC hierarchy adds avoidable complexity. If, however, the carriers belong to different RATs, legacy 5G/6G interworking arrangements, or backhaul conditions that make a common CA abstraction inefficient, then DC-like operation remains useful. Note that physical non-collocation should not by itself imply DC. An important intermediate deployment case is a distributed radio access network (RAN) in which multiple remote radios are connected to the same baseband through a lower-layer split and open fronthaul. In such a deployment, the antenna sites may be geographically separated, but the carriers can still belong to the same scheduling domain. CA may therefore be more appropriate than DC in such a distributed RAN. In short, the 6G baseline should therefore be CA-centric but not DC-blind.

From this perspective, enhanced CA should be the preferred baseline method for 6G spectrum aggregation. This does not mean that every deployment scenario must be forced into a tightly synchronized single-scheduler model. Rather, it means that 6G should first determine how far an enhanced CA framework can be extended before introducing a separate DC architecture. Recent 3GPP discussions already point in this direction: For non-collocated FR1 and FR2 aggregation with backhaul latency up to 10 ms, both 6G-6G DC with two cell groups and 6G-6G CA with two physical uplink control channel (PUCCH) groups and separate timing advance per group have been identified as capable of supporting imperfect-backhaul aggregation [15]. Importantly, the physical-layer functionalities of the two solutions are similar. This weakens the case for introducing DC as a parallel baseline mechanism for 6G-6G aggregation: if the same lower-layer functions can be provided by CA with two PUCCH groups and separate timing advance, then CA offers a cleaner path with less architectural duplication.

However, with large backhaul delay, the common-MAC abstraction in CA may become a throughput bottleneck. A centralized or logically common MAC scheduler needs sufficiently fresh information about channel quality, buffer status, HARQ state, and carrier load across all aggregated carriers. If a remote carrier is connected through backhaul with up to 10 ms delay, this information can become stale by the time it is used. The scheduler may then apply conservative link adaptation, under-schedule the remote carrier, or delay retransmission decision, resulting in throughput loss. DC avoids some of these MAC-layer penalties because each cell group has its own lower-layer protocol stack and can perform more local scheduling. However, DC moves the problem upward rather than eliminating it: split-bearer operation must handle PDCP reordering, path imbalance, inter-node flow control, and SCG lifecycle management. Thus, the 6G design choice should depend on where the transport delay is best absorbed. If delay can be hidden within a latency-aware CA framework, CA avoids unnecessary hierarchy. If delay would severely impair common-MAC scheduling, DC-like operation may provide better throughput by localizing MAC scheduling and pushing aggregation to PDCP or bearer-level mechanisms.

6G-6G CA with a single PUCCH group provides the most robust downlink-continuity behavior when the high-frequency uplink is coverage-limited. For example, an upper-mid-band or FR2 SCell may still provide useful downlink capacity even when the UE cannot reliably transmit PUCCH on that high-frequency layer. In single-PUCCH-group CA, HARQ-ACK and other uplink control information for the high-frequency downlink can be carried on a more robust lower-frequency uplink, such as an FR1 PCell uplink. This may introduce performance penalties, including higher uplink-control loading, less flexible scheduling, longer HARQ feedback timing, or additional backhaul-dependent latency in non-collocated deployments. Nevertheless, the high-frequency downlink carrier can remain usable.

6G-6G CA with two PUCCH groups provides a useful mode when both frequency layers have viable uplink-control coverage. Each group may maintain its own uplink-control resources and timing advance, allowing FR1 and upper-mid-band/FR2 legs to operate with more localized feedback timing and looser cross-carrier coordination. This is especially relevant for non-collocated deployments where the two transmission points do not share ideal backhaul or tight timing. However, two-PUCCH-group CA should not be viewed as universally superior to single-PUCCH-group CA. When the high-frequency uplink-control path is not sustainable, the single-PUCCH-group mode may be the only practical way to keep the high-frequency downlink carrier active while relying on a robust lower-frequency uplink for uplink control information (UCI).

6G-6G DC with two cell groups provides the loosest coordination model and remains useful when the two legs are sufficiently independent that CA assumptions become too restrictive. However, DC does not provide the same graceful downlink-only fallback as single-PUCCH-group CA. In DC, the SCG is not merely a downlink SCell added to the MCG. It has a PSCell and its own cell-group operation, which requires a sustainable uplink relationship for the SCG. If the UE cannot maintain the high-frequency SCG uplink, the network cannot simply keep using that SCG as a downlink-only CA carrier controlled by the MCG PUCCH. Instead, the SCG may need to be released or reconfigured, even if the high-frequency downlink radio channel still has useful capacity. Thus, single-PUCCH-group CA may suffer a throughput or latency penalty under weak high-frequency uplink coverage, but DC may lose the entire SCG. This is a key reason why 6G should improve CA-based control association rather than relying on DC as the default solution for FR1-FR2 or upper-mid-band aggregation. DC should be selected when independent cell-group operation brings a clear benefit, not simply because a high-frequency downlink carrier is physically separated from the lower-frequency coverage layer.

## IV. KEY ENHANCEMENT DIRECTIONS FOR 6G CA

To make CA suitable as the primary 6G spectrum-aggregation framework, 6G CA should address the main limitations observed in 5G CA and extend the framework. As illustrated in Figure 3, four enhancement areas are particularly important.

### *A. Latency-Aware Distributed CA*

5G CA already supports important ingredients for collocated and selected non-collocated deployments. In particular, multiple timing advance groups and PUCCH-group-based feedback handling provide mechanisms for handling different uplink timing and control assumptions within CA. The remaining problem is how efficiently CA works when distributed carriers are connected through non-ideal transport. In 5G, non-collocated CA can be deployed, but the common CA abstraction may impose time-critical dependencies between a remote downlink carrier and the carrier that manages uplink control. For example, in single-PUCCH-group operation, downlink transmissions from different locations may need to map their HARQ-ACK and other UCI to a common PUCCH resource. This preserves the coverage advantage of using a robust lower-frequency uplink, but it can add backhaul-dependent scheduling-to-action latency and reduce throughput.

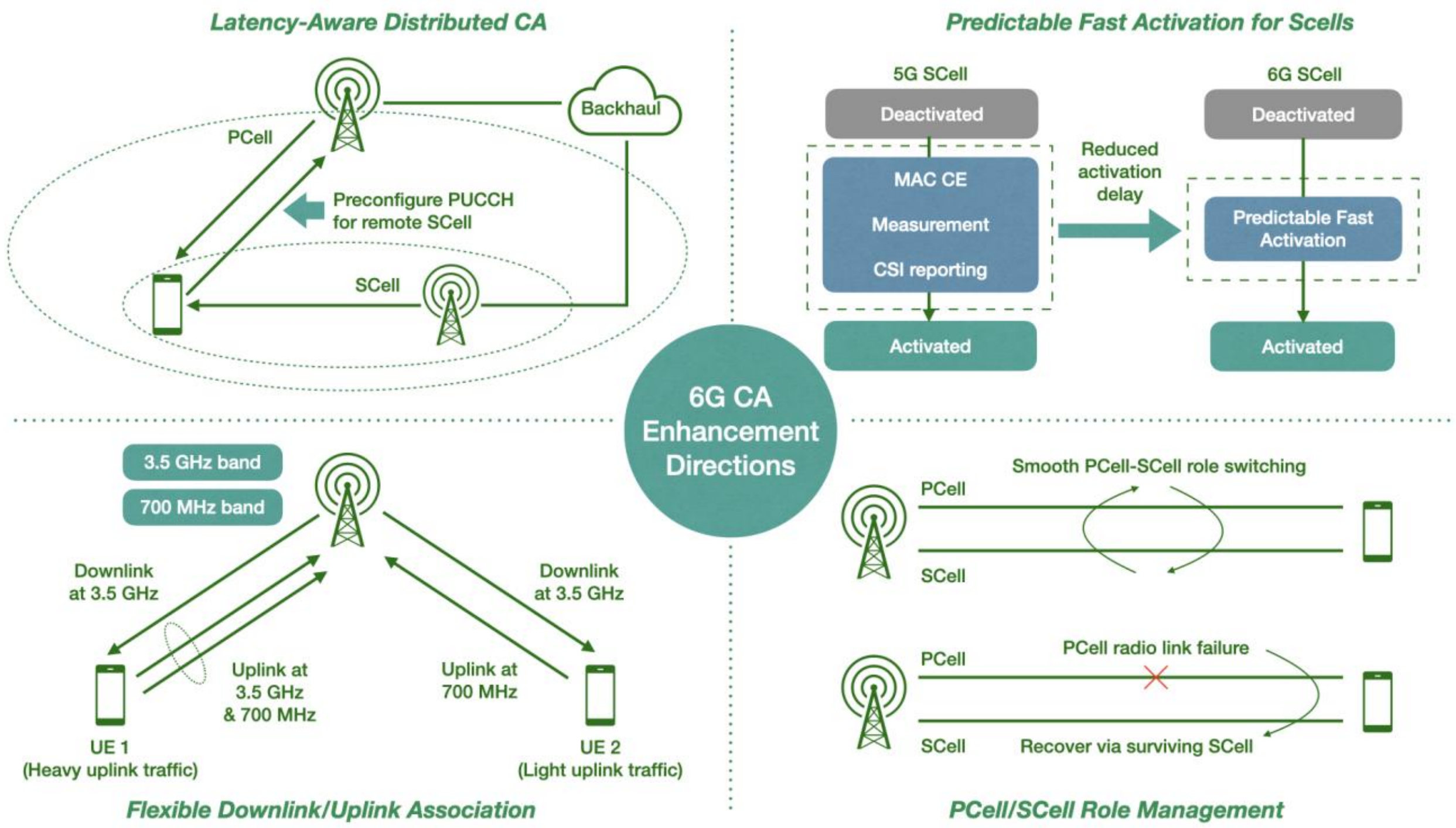

**Figure 3: An illustration of key enhancement directions for 6G CA.**

For single-PUCCH-group distributed CA, a key 6G enhancement is to avoid making every remote downlink scheduling decision depend on a real-time PUCCH-resource decision at the anchor carrier. Without pre-arrangement, a remote downlink scheduler may need to coordinate with the PUCCH-side scheduler before each physical downlink shared channel (PDSCH) transmission to determine whether the corresponding HARQ-ACK can be carried on the robust uplink carrier and which PUCCH resource should be used. If this coordination crosses non-ideal backhaul, the scheduling decision itself becomes backhaul-limited, not only the later HARQ-ACK forwarding. A better approach is to preconfigure PUCCH resource pools or feedback-resource patterns for a remote carrier group. The remote scheduler can then schedule PDSCH within a known feedback budget, while the UE maps the resulting HARQ-ACK to preconfigured PUCCH resources according to a configured rule. This converts a per-transmission backhaul coordination problem into a preconfigured feedback-mapping problem, preserving the coverage benefit of lower-frequency uplink control while reducing scheduling-to-action latency.

### B. *Predictable Fast Activation for SCells*

Reducing the latency of SCell activation is another important enhancement direction. One of the key shortcomings of 5G CA is that SCell activation can be too slow for bursty traffic. If a secondary carrier is activated only after the UE buffer has already drained, the additional carrier provides little user-perceived benefit. 5G already includes several mechanisms for reducing SCell activation latency and UE energy consumption, including fast SCell activation enhancements, SCell dormancy, and synchronization signal block (SSB)-less SCells when timing or measurement assumptions can be derived from another serving cell, among others. One remaining issue is that activation behavior is still not sufficiently predictable for bursty traffic and distributed carrier groups. A configured SCell may be known to the UE but have stale CSI, outdated beam information, invalid timing assumptions, or uncertain pathloss reference. Conversely, keeping many SCells fully active to avoid this uncertainty increases PDCCH monitoring, CSI measurement, RF chain activity, and baseband power consumption.

One potential 6G enhancement is to standardize a fast-activation eligibility profile for an SCell or SCell group. The profile would specify a reference serving cell, the context that may be inherited from that reference cell, the validity condition for the inherited context, and an activation-time class. The inherited context may include timing reference, pathloss reference, beam relation, RRM validity, or CSI freshness. The validity condition may be based on a timer, collocation or intra-band assumptions, measurement freshness, mobility state, or beam-relation validity. If the profile remains valid, the network may activate and schedule the SCell group after the configured activation offset. If the profile is invalid, the UE falls back to normal activation or waits for additional reference signals or measurements. This approach turns existing 5G mechanisms into a predictable carrier activation scheme: carriers do not need to remain fully active, but the network knows when they are safe to use quickly.

### C. *Flexible Downlink/Uplink Association*

In many deployments, the best downlink carrier will not be the best uplink carrier. For example, upper-mid-band or FR2

carriers may provide large downlink bandwidth, while low-band or lower-mid-band carriers may provide better uplink coverage, more reliable uplink control, and lower UE power burden. A rigid assumption that downlink and uplink should remain paired on the same carrier would limit the value of multi-band aggregation. 5G already introduced the basic principle of flexible downlink/uplink association through supplementary uplink (SUL), where a UE may use an additional, often lower-frequency, uplink to improve coverage while downlink service is provided on another carrier. However, SUL has limited deployment in 5G. It requires suitable low-frequency uplink spectrum, UE support for the relevant band combination, RF-front-end coexistence support, and careful planning of physical random access channel (PRACH), PUCCH, physical uplink shared channel (PUSCH), timing, etc., across the normal uplink and SUL. Moreover, SUL is not a general flexible association framework. The UE is scheduled on either the normal uplink or the SUL, overlapping PUSCH/PUCCH transmissions on the two uplinks are avoided, and PUCCH is configured on only one of the two uplinks of the cell.

The 6G lesson is therefore not that SUL should simply be repeated, but that its useful principle should be generalized into a CA framework for deciding how a multi-carrier downlink group should obtain reliable uplink control while using uplink data resources opportunistically. One potential extension of SUL is an uplink association set for CA. Instead of configuring one normal uplink and one SUL with essentially exclusive use, the network would configure a small set of uplink carriers for a serving cell or carrier group and assign them roles. One uplink carrier may serve as the primary UCI anchor for HARQ-ACK, scheduling request, and critical CSI; an optional secondary UCI anchor may provide fallback or load sharing; and one or more uplink carriers may be used for uplink data. The UE capability would indicate whether these uplink carriers are used exclusively or whether controlled overlap is supported. The standard may define priority, power-sharing, and fallback rules so that critical UCI is protected while uplink data remains opportunistic. This relaxes the main limits of SUL and provides a carrier-group framework in which high-frequency downlink capacity can rely on robust lower-frequency uplink control, while uplink data is scheduled on one or more carriers that best match coverage, power headroom, capacity, and UE capability.

### *D. PCell/SCell Role Management*

Another important enhancement direction is to relax the rigid PCell-centric behavior of 5G CA. In 5G, a PCell change is treated as a handover, even when the desired operation is essentially a role change between already configured serving cells. This can introduce interruption and signaling overhead that is disproportionate to the intended change. Similarly, when PCell radio link failure occurs, RRC re-establishment is triggered and the configured SCells are released, even if some secondary carriers remain usable and their configuration context is still valuable. These behaviors are inefficient for a 6G system that may rely on multiple carriers for coverage, capacity, control, and resilience.

For 6G CA, the enhancement should therefore be CA-context-preserving role management. The network could preconfigure selected SCells as anchor-capable candidates within a prepared multi-carrier context. Such candidates would typically be coverage-suitable carriers with the necessary control, timing, and measurement context, rather than opportunistic capacity-only carriers. When the network needs to change the serving anchor, the UE and base station could perform a lightweight PCell/SCell role transfer within the configured CA context instead of always executing a full handover-like procedure. Failure recovery should also become carrier-group-aware. If the current PCell fails but another prepared carrier remains reliable, the UE should be able to use that surviving carrier to assist recovery, preserve applicable parts of the CA configuration, and avoid unnecessary teardown of all SCells.

This enhancement complements, rather than duplicates, L1/L2-triggered mobility. L1/L2-triggered mobility reduces interruption for cell switching, while CA-context-preserving role management targets a different inefficiency: the unnecessary loss of multi-carrier aggregation state when the PCell role changes or the PCell fails. The benefit for 6G is lower interruption, lower signaling overhead, and more resilient CA operation in multi-carrier deployments where different carriers may have different coverage, interference, and reliability characteristics.

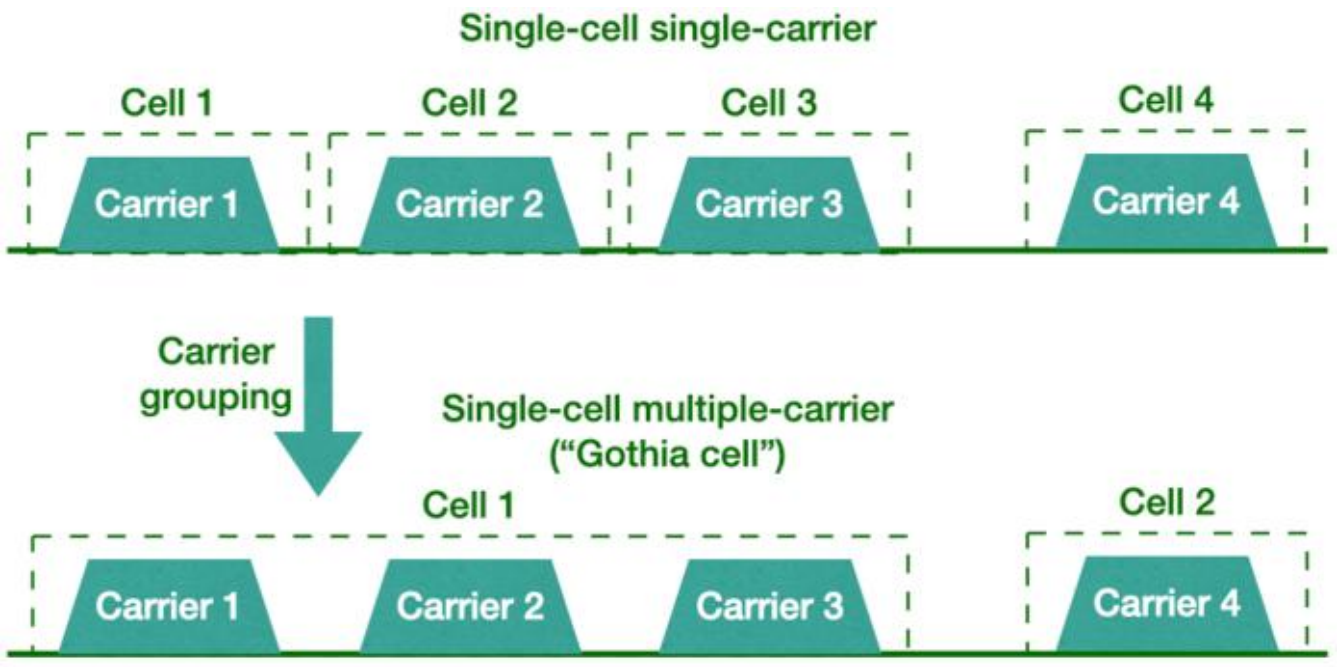

**Figure 4: An illustration of carrier grouping for fragmented-carrier operation.**

## V. EFFICIENT FRAGMENTED-CARRIER OPERATION

The preceding sections focused on how 6G can improve CA as the scalable carrier-level aggregation baseline and how DC-like operation should remain available when independent cell groups are genuinely needed. A related issue, currently being discussed in 3GPP, is how 6G should operate efficiently over fragmented narrow carriers, as illustrated in Figure 4. The key objective is that when multiple narrow carriers are close in frequency and share suitable deployment properties, the system should avoid duplicating synchronization, measurement, control, activation, and mobility overhead for every fragment.

This objective has important 5G precedents. NR already supports non-contiguous CA, and several mechanisms reduce per-SCell overhead when the relationship between serving cells is sufficiently strong. For example, SSB-less SCells can be supported in suitable cases where timing or measurement assumptions can be derived from another serving cell. Similarly, RRM measurement and mobility treatment can be simplified for some collocated or intra-band SCells. Therefore, the question for 6G is how to make fragmented-carrier operation systematic across a wider range of deployments. One potential 6G direction is a fragmented-carrier set profile. Such a profile would define when multiple narrow carriers may be treated as a coordinated set for selected procedures. The profile could specify a reference carrier for timing, pathloss, or measurement assumptions; conditions for reduced or SSB-less operation on secondary fragments; simplified RRM and mobility treatment; common activation or dormancy handling; and possibly compact control signaling for scheduling across the set. The key is to specify the assumptions under which per-fragment overhead can safely be reduced. This complements the CA-centric design discussed above: enhanced CA provides the general carrier-scaling framework, while fragmented-carrier profiles reduce avoidable overhead when the carriers are sufficiently aligned to be managed as a coordinated set.

| Aspect | 5G Carrier Aggregation | 5G Dual Connectivity | Implication |
|---|---|---|---|
| **Architectural role** | Carrier-level aggregation within a serving-cell framework. | Higher-layer aggregation across MCG and SCG. | CA and DC are hierarchical; DC may contain CA in each cell group. |
| **Scaling behavior** | Adds DL/UL carriers, often with fewer UL/control carriers than DL carriers. | Adds a second cell group, each requiring viable uplink/control. | CA is the scalable carrier layer; DC should not be used merely for carrier scaling. |
| **Weak high-frequency UL** | DL CA can continue using lower-frequency UCI, with latency or control-loading penalty. | SCG may need release if the SCG uplink/control path is not sustainable. | Improve single-PUCCH-group CA and UCI anchoring. |
| **Scheduler and transport** | One MAC framework can jointly schedule carriers, but may suffer stale information under large backhaul delay. | Local MAC scheduling in each cell group, but PDCP/inter-node complexity. | Use latency-aware CA when transport is bounded; use DC when common-MAC CA loses efficiency. |
| **Deployment topology** | Co-sited, distributed RAN, and selected non-collocated deployments. | Different RATs, independent schedulers, or large transport delay. | Physical non-collocation alone does not imply DC. |
| **Traffic time scale** | Preconfigured/dormant SCells can serve bursts within one carrier-level framework. | SCG setup may complete after short bursts have already drained. | Compare flow-completion time and setup/activation overhead, not only peak rate. |
| **Bearer complexity** | Carrier multiplexing within one aggregation framework. | MCG/SCG/split bearers, PDCP routing, reordering, flow control, duplication. | Use split bearers only when independent paths justify the complexity. |
| **Uplink power sharing** | One scheduler can coordinate UE power headroom and uplink power sharing across carriers. | MCG/SCG schedulers may lack timely knowledge of each other's UL power usage. | CA is cleaner for power-aware UL aggregation when one scheduling domain is feasible. |
| **Mobility/failure** | PCell-centric behavior can cause unnecessary role-change and recovery overhead. | SCG addition/release, PSCell change, and SCG failure procedures add flexibility and complexity. | 6G CA should preserve aggregation context during PCell/SCell role changes and recovery. |
| **6G role** | Preferred baseline for coordinated multi-band spectrum aggregation. | Reserved for legacy integration, or very non-ideal transport. | 6G should be CA-centric but retain DC where a second cell group is genuinely needed. |

**Table 2: 5G CA versus DC: technical differences and implications.**

UE capability is also critical. Fragmented spectrum is attractive in lower bands, including sub-GHz deployments where coverage and IoT support are important. However, reduced-capability and IoT UEs may have limited maximum RF bandwidth, simplified receiver design, and tighter energy constraints. Such UEs may not be able to monitor or process a wide fragmented set without per-carrier synchronization or measurement support. Therefore, 6G should allow fragmented-carrier profiles to be UE-capability-dependent. High-end UEs may benefit from shared reference carriers, compact control, and reduced RRM overhead across fragments, while constrained UEs may require conventional SSB and RRM treatment per carrier.

## VI. CONCLUSIONS

Spectrum aggregation will remain a foundational capability for 6G as networks combine fragmented spectrum across different frequencies. Table 2 summarizes the main differences between 5G CA and DC and the corresponding implications discussed in this article. The main lesson from 5G is that CA and DC are hierarchical rather than interchangeable. CA is a scalable carrier-level mechanism, while DC is a higher-layer cell-group architecture that may itself contain CA within each cell group. This article has argued for a CA-centric 6G aggregation baseline. Enhanced CA should be the default framework when carriers can be coordinated with acceptable timing, control, scheduling, and transport assumptions, while DC-like operation should remain available when legacy integration or large transport delay justifies a second cell group. Together with efficient fragmented-carrier operation, this approach can help transform diverse spectrum assets into a scalable and deployment-flexible 6G radio access system.